\documentclass[conference]{IEEEtran}
\IEEEoverridecommandlockouts
\usepackage{amsmath,amssymb,amsfonts}
\usepackage{algorithmic}
\usepackage{graphicx}
\usepackage{textcomp}
\usepackage{xcolor}
\usepackage{pdflscape}
\usepackage{comment}
\usepackage{acro}
\usepackage{float}
\usepackage{multirow}
\usepackage{nicematrix}
\usepackage{lipsum}
\usepackage[table]{xcolor}
\usepackage[dvipsnames]{xcolor}
\usepackage{color,soul}
\usepackage{tikz}
\usepackage{stmaryrd}
\usepackage[flushleft]{threeparttable}
\usetikzlibrary{arrows.meta}
\usetikzlibrary{fit}
\usepackage[table]{xcolor}
\usepackage[dvipsnames]{xcolor}
\usepackage{color,soul}
\usepackage{caption}
\usepackage{subcaption}
\usepackage{array}   
\definecolor{lightyellow}{RGB}{255, 242, 204}
\definecolor{lightgreen}{RGB}{217, 234, 211}
\definecolor{lightblue}{RGB}{173, 216, 230}
\definecolor{lightred}{RGB}{244, 204, 204}
\usepackage[
    left=0.5in,
    right=0.5in,
    top=0.5in,
    bottom=0.5in
]{geometry}

\def\BibTeX{{\rm B\kern-.05em{\sc i\kern-.025em b}\kern-.08em
    T\kern-.1667em\lower.7ex\hbox{E}\kern-.125emX}}

\DeclareAcronym{sdo}{
  short=SDOs,
  long=Standards Developing Organizations,
}
\DeclareAcronym{ai}{
  short=AI,
  long=Artificial Intelligence,
}

\DeclareAcronym{xr}{
  short=XR,
  long=Extended Reality,
}

\DeclareAcronym{vr}{
  short=VR,
  long=Virtual Reality,
}

\DeclareAcronym{3g}{
    short=3G,
    long=Third Generation Cellular Networks,
}

\DeclareAcronym{ues}{
    short=UEs,
    long=User Equipments,
}

\DeclareAcronym{ran}{
    short=RAN,
    long=Radio Access Network,
}

\DeclareAcronym{cn}{
    short=CN,
    long=Core Network,
}

\DeclareAcronym{iot}{
    short=IoT,
    long=Internet of Things,
}

\DeclareAcronym{gdp}{
    short=GDP,
    long=Gross Domestic Product,
}

\begin{document}
%On the Mathematical Formulation of Communication-Control Co-Design in Mobile Broadband Systems
%\title{On the Mathematical Formulation of Communication-Control Co-Design Problems in Mobile Broadband Systems} 
\title{A Compositional Perspective on Communication-Control Co-Design for Mobile Broadband Systems Beyond 6G}

\author{Angelo Vera-Rivera,~\IEEEmembership{Member,~IEEE}, Atefeh Termehchi, and Ekram Hossain,~\IEEEmembership{Fellow,~IEEE}.  

\thanks{The authors are with the Department of Electrical and Computer Engineering, University of Manitoba, Winnipeg, Manitoba, Canada (e-mail: verariva@myumanitoba.ca, ekram.hossain@umanitoba.ca, and atefeh.termehchi@umanitoba.ca).}}

\maketitle
%\textcolor{green}{\textit{The abstract has been reorganized to clearly reflect the motivation and contribution of our work.}}\\, such as advanced immersive environments and networked physical regulation,become less intuitive to develop as the number of interacting subsystems increases.
\begin{abstract}
Anticipated applications of beyond sixth-generation (B6G) mobile broadband networks will require the co-design of communication and control subsystems within the network architecture. Existing co-design approaches are predominantly optimization-driven, integrating subsystems through joint optimization problems. While this approach is effective in relatively simple systems, such formulations present challenges in (i) modular subsystem representations, (ii) tracing the propagation of requirements across subsystems, and (iii) systematically analyzing design-space feasibility, particularly as the number and complexity of interacting subsystems increase. In this article, we present a compositional perspective on co-design based on the formal theory of co-design. We examine how the notion of composition introduced by this theory can address the challenges of the conventional optimization-driven approach. Building on this perspective, we propose a composition-driven methodology for communication-control co-design in B6G networks and illustrate it through a wireless-assisted robotic control case-study. We then discuss how the composition-driven and optimization-driven co-design approaches can complement each other and why this complementarity may be beneficial for B6G networks. Finally, we identify key research challenges and future directions toward the practical adoption of the proposed methodology. 
\end{abstract}
\begin{IEEEkeywords}
Beyond sixth-generation (B6G), communication-control co-design, composition-driven co-design, formal theory of co-design.
\end{IEEEkeywords}

%Context
%Trends and Motivation
%Research Gap
%Contributions
%Organization
\section*{Introduction: From Data Delivery to Cyber-Physical Connectivity in B6G}
Since the early 2000s, mobile broadband systems have been a primary enabler of wireless Internet access. Across successive generations, their design has remained largely communication-centric, with network guarantees focused on information exchange. Although IMT-2030\footnote{The IMT-2030 framework is the technical framework for sixth-generation (6G) cellular networks defined by the International Telecommunication Union (ITU).} introduces capabilities beyond data delivery, the underlying architecture remains largely optimized for information transfer, where network primitives are expressed through communication key performance indicators (KPIs)~\cite{ITU2023}.

In recent years, accelerated by the COVID-19 pandemic and the widespread adoption of telepresence, academia and industry leaders have increasingly envisioned the Internet as a platform for (i) advanced immersive experiences and (ii) networked physical regulation~\cite{lee2024all}. The former refers to \textit{human-centric}, multisensory environments that seamlessly integrate the physical and digital worlds. The latter involves the monitoring and control of \textit{machine-centric} physical systems over communication networks. Both require communication and control subsystems to maintain desired operating conditions. 

%capable of synchronizing physical and digital realities, as well as enabling physical regulation.
To support these emerging Internet trends, we argue that future mobile broadband beyond sixth-generation (B6G) must evolve from purely data delivery toward systems with cyber-physical capabilities. In systems engineering, cyber-physical systems (CPSs) are engineered systems that integrate communication, computation, control, and physical processes into closed-loop architectures.\footnote{In control theory, a closed loop refers to a feedback loop in which system outputs are measured and used to adjust subsequent control inputs in order to achieve a desired behavior.} Such architectures enable computing and communication infrastructures to monitor the state of physical systems and generate control actions that influence their dynamics~\cite{Wiener1948Cybernetics}. 

CPSs integrate heterogeneous subsystems, including communication and control, to regulate physical processes. Their interaction is commonly addressed through communication-control co-design, typically formulated as a joint optimization problem. In conventional optimization-driven co-design, subsystem models are embedded directly within this formulation and design choices are obtained from its solution. As system complexity grows, however, this approach faces three key challenges: (i) subsystem models become highly embedded within a joint formulation, limiting modularity and making extensions or reconfiguration difficult; (ii) the propagation of functionality requirements and resource demands across interacting subsystems becomes difficult to trace explicitly; and (iii) feasibility analysis and final design selection are handled simultaneously, providing limited visibility into the feasible design space and often requiring repeated redesign when infeasibility occurs.

In this article, we present a complementary co-design approach based on the notion of composition introduced by the formal theory of co-design. Unlike conventional optimization-driven approaches, the compositional perspective provides theoretical tools for modular subsystem representation through a common abstraction, explicit representation of functionality-resource dependencies through formal composition rules, and the characterization of Pareto-optimal design alternatives through functionality-resource feasibility analysis of the composed system.

We argue that communication-control problems in future cyber-physical B6G networks may benefit from the complementary use of both co-design approaches, as their architectures are expected to become increasingly heterogeneous and highly interconnected, involving subsystems with different functionalities and resource requirements. Our main contributions are:
\begin{enumerate}
    \item A technical discussion motivating communication-control co-design in cyber-physical B6G and contrasting conventional optimization-driven and composition-driven approaches;
    \item A composition-driven methodology for communication–control co-design, illustrated through a wireless-assisted robotic control case study;
    \item A discussion of research challenges and future directions.
\end{enumerate}
%%    \item An illustrative case study applying the proposed methodology to a wireless-assisted robotic control system; and
The rest of the article is organized as follows. We first discuss the need for communication-control co-design in cyber-physical B6G and contrast conventional optimization-driven and composition-driven approaches. We then introduce the proposed methodology. Finally, we discuss research challenges and future directions before concluding the article. 
\section*{On the Need for Communication-Control Co-Design in Cyber-Physical B6G}
Control is the process of designing and applying inputs to a dynamical system to regulate its behavior and achieve desired objectives~\cite{AstromMurray2008}. These inputs are determined by a controller that implements a given control strategy. In closed-loop control, measurements of the system state are fed back to the controller to determine the control inputs. When the dynamical system and the controller are physically separated, a communication network (e.g., mobile broadband) is required to establish forward and feedback paths between them. For systems with relatively slow dynamics, where the system evolves on a timescale much longer than the communication delays, communication performance typically has limited impact on the closed-loop behavior. In these systems, the controller and communication network can often be designed independently. The situation changes fundamentally when the dynamical system evolves on a timescale comparable to communication delays, making the communication and control designs mutually dependent. 
%Examples include advanced immersive experiences requiring real-time synchronization between physical and digital worlds and autonomous machine-type systems whose stability depends on strict control-loop timing requirements.

The recognition of dependencies between communication and control in mobile broadband is not new. In their seminal 2019 vision paper on 6G, Walid Saad \textit{et al.}~\cite{8869705} highlighted the limitations of communication-centric architectures for supporting connected robotics and autonomous systems (CRAS),\footnote{CRAS include autonomous robots, drone swarms, and autonomous vehicles.} a key use case for 6G and beyond. They noted that \textit{``[...] the past five generations of cellular systems had one exclusive function: wireless communications. However, 6G will disrupt this premise through a convergence (i.e., joint and simultaneous offering) of various functions that include communications, computing, control, localization, and sensing (3CLS)."} They argued that, in CRAS, the physical environment becomes part of a control system and that 3CLS services are essential to its operation. Despite this early insight, control capabilities were not explicitly incorporated into IMT-2030. 
%realizing them. An explicit recognition that control functions are inherent of these architectures. 
%They continue, \textit{``[...] the introduction of CRAS over the cellular domain is not a simple case of `yet another short-packet uplink Internet of Everything (IoE) service.' [...] Instead, the physical environment becomes a control system, potentially augmented with AI.”} Despite this early insight, control functions were not explicitly incorporated into IMT-2030. 

\begin{figure*}[htbp]
\centering
\begin{subfigure}[b]{\textwidth}
{
\centering
\includegraphics[width=0.69\textwidth]{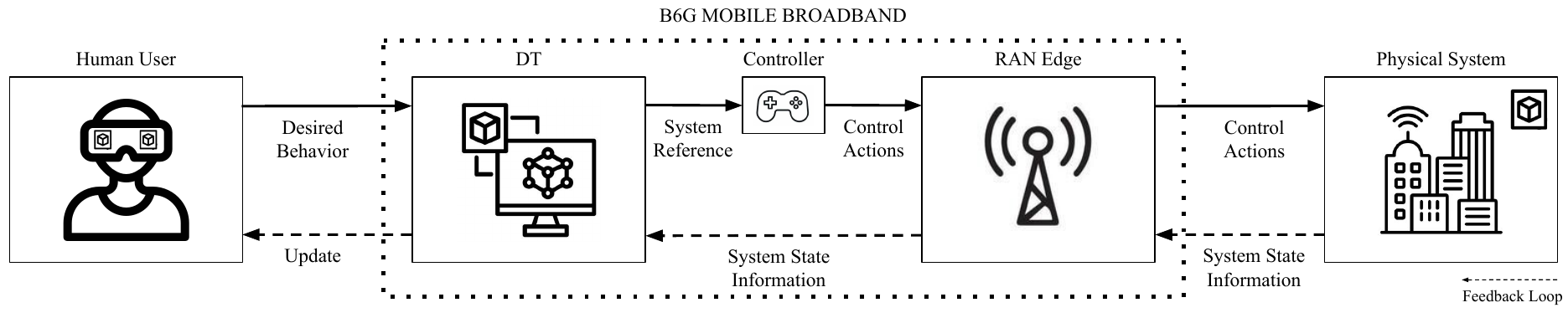}
\caption{Human-centric CPIE: Human users interact with the physical system through synchronized DTs in a closed feedback loop. Users specify the desired system behavior, while the physical system continuously updates the DTs with its current state.}
\label{fig:CPIE}
}
\end{subfigure}
%\hfill
%\bigskip
%\vspace{2cm}
\begin{subfigure}[b]{\textwidth}
{
\centering
\includegraphics[width=0.69\textwidth]{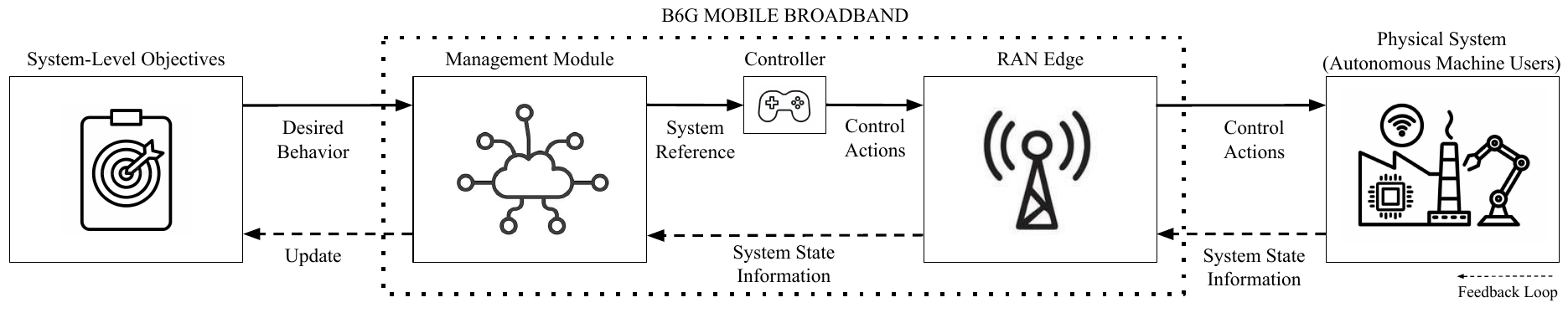}
\caption{Machine-centric ISCAC: The network continuously monitors the autonomous machines. The resulting state information is fed back to RAN-edge controllers, which generate the corresponding control actions according to system-level objectives.}
\label{fig:ISCAC}
}
\end{subfigure}
\caption{Representative B6G usage scenarios requiring native control capabilities~\cite{hossain2025hyperdimensionalconnectivity6gconceptual}.}
\label{fig:useCasesB6G}
\end{figure*}

%, rather than information transfer alone, The framework defines connectivity across four dimensions: communication, cognition, computation, and cyber-physical regulation. 
Inspired by Saad's CRAS paradigm, in 2026 we proposed a conceptual connectivity framework for B6G that moves beyond communication-centric architectures~\cite{hossain2025hyperdimensionalconnectivity6gconceptual}. Within this framework, physical-state regulation is a primary network objective and native control subsystems become fundamental architectural components. Specifically, we identify two representative B6G usage scenarios requiring native control capabilities: (i) \textit{Cyber-Physical Immersive Environments (CPIE)} and (ii) \textit{Integrated Sensing, Communication and Control (ISCAC)}. Figure~\ref{fig:useCasesB6G} illustrates the proposed scenarios. 
%

%~\cite{Tao2019DigitalTwin}
CPIE evolves from IMT-2030’s immersive communication usage case by allowing users to interact with synchronized physical and digital worlds. We refer to this scenario as advanced immersive experiences. Digital twins (DTs)\footnote{DTs are digital replicas of physical assets, processes, and systems that are continuously updated using real-time data and capable of mirroring, simulating, and predicting the behavior of their physical counterparts.} are central to CPIE, serving as the cyber interface to the physical world. Users interact with DTs located at the RAN edge to define the desired behavior of the physical system. As the system evolves, sensing and communication continuously update the DTs with its state information, forming a closed-loop cyber-physical architecture. In the DT-to-physical direction, changes in the DTs translate into control actions sent to the physical system. In the physical-to-DT direction, sensing and communication deliver the system state back to the DTs to maintain an accurate digital representation.

ISCAC extends IMT-2030’s integrated sensing and communication (ISAC) by incorporating native control subsystems that establish closed-loop architectures between the RAN edge and autonomous machine-type systems. In conventional ISAC, the network primarily functions as an observation and information-delivery system. It acquires contextual information for situational awareness (e.g., localization, tracking, occupancy, and gesture recognition) and exposes it to higher-layer applications for decision-making and actuation. In ISCAC, sensed information instead serves as feedback for RAN-native controllers located at the network edge. These controllers generate control actions transmitted back to the autonomous system through actuators. As the physical system evolves, its dynamics alter the timing requirements of the control loop. Accordingly, the RAN reconfigures its communication resources to accommodate these changes. 
%Applications with relatively slow dynamics can tolerate infrequent control updates and moderate communication delays. In contrast, fast dynamics require frequent control updates with strict latency and reliability guarantees to maintain stable operation. %ISCAC closes this gap by embedding control directly within the mobile broadband network.

Both CPIE and ISCAC involve highly dynamic systems in which closed-loop control is not merely an auxiliary function but a fundamental component of the mobile broadband architecture. In these systems, communication and control become mutually dependent: control objectives determine the communication requirements, while communication performance limits the achievable control objectives. If designed independently, delayed, unreliable, or inconsistent information exchange directly affects control objectives. Consistent with the vision of Saad \textit{et al.}, we call for co-design approaches that appropriately capture these interactions and fundamental tradeoffs. 
%we argue that communication and control must be designed jointly within B6G architectures.  We therefore 
%
\section*{Optimization-Driven versus Composition-Driven Co-Design}
%This approach addresses general optimization problems with multiple sets of disciplinary design variables.
%In control theory, control co-design (CCD) is a well-established design strategy for systems requiring embedded control mechanisms~\cite{https://doi.org/10.1002/adc2.18}. CCD can be viewed as an instance of MDO in which the control system is coupled with the remaining subsystems through joint optimization. 
Multi-disciplinary optimization (MDO) is an engineering approach based on optimization methods to solve design problems involving multiple interacting subsystems. In the literature, this approach is largely referred to as co-design~\cite{https://doi.org/10.1002/adc2.18}. Conventionally, designers first design problem-specific subsystem models, objective functions, and constraints, and then formulate a joint optimization problem that guides both the integration analysis and the resulting design decisions. These problems are commonly formulated in one of two ways: (i) one domain is optimized and the other is treated as a constraint, or (ii) both domains are optimized simultaneously under shared constraints~\cite{8166737}.
%CCD has been successfully applied across numerous engineering domains, including aerospace, mechanical engineering, power systems, robotics, and wireless communications. In wireless communications literature, optimization-driven co-design is 

For instance, the authors in~\cite{8761867} adopt a control-objective formulation of approach (i) stated above, proposing a control solution whose actions adapt to the underlying communication conditions. They minimize a control cost function defined in terms of the control-input magnitude and the system-state deviations from the desired operating points. The problem is constrained by communication variables such as bandwidth, delay, and packet loss probability. By contrast, the authors in~\cite{8767991} adopt a communication-objective formulation of approach (i), designing an ultra-reliable low-latency communication (URLLC) service for wireless control systems. They maximize uplink spectral efficiency through the allocation of communication resources subject to control convergence constraints. Under approach (ii), the authors in~\cite{11186883} propose an uncertainty-aware task offloading model for safe trajectory generation in autonomous industrial robots. They jointly optimize the communication rate and robot speed under a location uncertainty constraint, limiting the amount of data offloaded to the network edge while maintaining safe robot operation and reducing network congestion and edge-computing demand. Regardless of whether approach (i) or (ii) is adopted, the resulting formulations present several challenges discussed next.

% designed on a case-by-case basis. %Poor Modularity in Large Co-Design Formulations with Complex Interactions Among Subsystems}ad-hoc basis\footnote{In engineering, an ad hoc approach typically refers to a method or solution designed for a specific problem or purpose, rather than for general use.}
\subsubsection*{Challenge 1--Co-Design Formulations May Be Difficult to Represent Modularly} In this conventional co-design approach, subsystem abstractions are embedded directly within a single joint optimization problem, with performance objectives, decision variables, and constraints defined on a problem-specific basis. While this approach is effective for systems with relatively simple structures, we argue that it becomes challenging to apply as the number and complexity of interacting subsystems grow. As more subsystems interact, their models become increasingly coupled, making individual subsystems difficult to modularly extend, reconfigure, or replace without reconstructing the co-design formulation. 
%Consequently, this approach limits the modular representation of subsystems and their structural dependencies, particularly in large systems. 
%
\subsubsection*{Challenge 2--The Propagation of Subsystem Requirements May Be Difficult to Trace}
In these formulations, system requirements and interactions among subsystems are expressed through shared objective functions, decision variables, and constraints. While these elements enforce subsystem dependencies, it may be difficult to determine how requirements imposed on one subsystem propagate into requirements and resource demands in others, particularly in formulations involving multiple interacting subsystems. 
%however, in large formulations with several interacting subsystems, it may be difficult to determine how requirements in different subsystems propagate across others. As a consequence, in this approach the dependencies among subsystems may be obscured, making it difficult to trace how subsystem requirements propagate across the co-design formulation. 
%In sum, tracing how  becomes increasingly difficult in large systems. For example a control-performance requirement may induce communication requirements such as latency or data rate, which in turn determine the consumption of wireless resources such as bandwidth and transmit power.translate into requirements and resource demands in

\subsubsection*{Challenge 3--Feasibility and Design Selection May Be Difficult to Separate} In this approach, the feasibility analysis of the system design space and the final design selection are normally handled by the same formulation. Solving the optimization problem determines whether a feasible solution exists and, if so, identifies an optimal design according to the specified objective. When the problem is infeasible, designers usually modify constraints through trial and error and repeatedly solve the problem until feasibility is achieved, which may be computationally inefficient for large systems. 

In the following sections, we will present a {\em composition-driven co-design approach}  (e.g., for designing communication and control subsystems) for cyber-physical B6G networks. 
%. As a consequence, this approach may be unstructured and computationally inefficient for large systems. The designer may run the optimization somewhat blindly. 
%
\subsection*{A Compositional Perspective on Co-Design}
\begin{table*}[t]
\centering
\scriptsize
%\footnotesize
\caption{Comparison between optimization-driven and composition-driven co-design. The compositional approach introduces explicit DPI-based subsystem abstractions, structured composition rules, and systematic characterization of functionality-resource tradeoffs.}
\label{tab:design_comparison}
\resizebox{\textwidth}{!}{
\begin{threeparttable}
%\begin{tabular}{|l|l|l|l|}
\begin{tabular}{|p{4.5cm}|p{4.5cm}|p{4.5cm}|}
\hline
\textbf{Dimension} & \textbf{Optimization-Driven Co-Design} & \textbf{Composition-Driven Co-Design} \\
\hline
\textbf{Mathematical Foundation} & Optimization theory & Category theory \\
\textbf{Engineering Design Approach} & MDO & Formal theory of co-design \\
\textbf{Subsystem Abstraction} & Problem-specific & DPI-based \\
\textbf{Integration Strategy} & Joint optimization formulation & Formal DPI composition \\
%\textbf{Composition Approach} & None &  & \\
%\textbf{Modularity} & Embedded in joint formulations & DPI abstractions + composition rules \\
\textbf{Feasibility Analysis} & Constraint-based feasibility & Functionality-resource feasibility maps \\
%\textbf{Tradeoff Analysis} & Multi-objective analysis & Resource Pareto fronts  \\
\textbf{Design Selection} & Optimization-based selection & Pareto-informed selection* \\
\hline
\end{tabular}
\begin{tablenotes}
    \item *May involve a secondary optimization over the Pareto-optimal alternatives.
  \end{tablenotes}
\end{threeparttable}
}
\end{table*}
%In practice, engineers naturally design complex systems
In several engineering fields, such as robotics, complex systems are viewed as the assembly or composition of interacting subsystems, each consuming resources to provide specific functionalities. For example, a robotic system may be composed of a chassis, energy supply, sensors, communication links, software, and processing units. The chassis uses structural materials to provide mechanical support; the energy supply converts external energy (e.g., solar or chemical) into power; communication links use bandwidth and transmission power to provide connectivity; and sensors consume energy to provide environmental measurements.  

However, subsystem designs are frequently not independent, as design choices in one subsystem may change the functionality or resource requirements of others. For instance, a more sophisticated sensing algorithm may reduce communication overhead while requiring higher-quality sensors and greater computational resources. The resulting increase in energy consumption may require larger batteries and a stronger chassis. The notion of composition motivates a systematic way to represent subsystems through their functionality and resource interfaces and explicitly represent the dependencies among them.
%\textcolor{blue}{My suggestion: Such subsystems are not independent. A design choice in one subsystem may change the functionality or resource requirements of others. For example, a more sophisticated sensing algorithm may reduce communication overhead while requiring higher-quality sensors and greater computational resources, which may in turn increase energy consumption and impose additional requirements on the battery and chassis. This motivates a systematic way to represent subsystem interfaces and formally describe how their functionality-resource relationships are interconnected.}

\textbf{Category theory} is a branch of mathematics that provides a framework for representing objects, their relationships, and the rules governing their composition~\cite{maclane:71}. The formal theory of co-design adopts these compositional principles to model engineering design problems through common functionality-resource interfaces and their interconnections~\cite{Fong_Spivak_2019}. It introduces two main theoretical tools: (i) a modeling tool for the abstraction and formal composition of interacting design problems, and (ii) an analysis tool for characterizing the functionality-resource dependencies of the resulting composed systems. Next, we discuss how these compositional tools can address the challenges of the optimization-driven co-design approach. 
%\textcolor{blue}{Category theory provides a mathematical framework for representing objects together with their relationships and rules of composition~\cite{maclane:71}. The formal theory of co-design adopts these compositional principles to model engineering subsystems through common functionality-resource interfaces and to combine them while preserving their interdependencies~\cite{Fong_Spivak_2019}. It provides two main theoretical tools: (i) a modeling tool for abstracting and formally composing interacting design problems, and (ii) an analysis tool for characterizing the functionality-resource relationships of the resulting composed system. Next, we discuss how these tools address the limitations of conventional optimization-driven co-design.}
%
\subsubsection*{Addressing Challenge 1---Modular Subsystem Representation Based on a Common Abstraction} The modeling tool represents each subsystem independently using a mathematical abstraction known as a design problem with implementation (DPI). Rather than exposing all internal equations and implementation details, a DPI describes the subsystem through two external interfaces: the functionality it provides and the resources it consumes. In this way, different subsystem models can be treated as modular building blocks, much like LEGO blocks. These blocks can then be formally interconnected through their functionality-resource interfaces to construct a representation of the larger system.
%composition-driven co-design approach represents subsystems independently using the DPI abstraction. A DPI provide a modular representation for subsystems that encapsulate their internal models and only expose their well-defined functionality and resource interfaces. As a consequence, subsystems can be treated as modules that can be "assembled" to form a larger system. This modularity allows subsystems to be developed, modified, or replaced individually without reconstructing the complete co-design representation. 

% according to the system architecture
\subsubsection*{Addressing Challenge 2---Explicit Representation of Inter-Subsystem Dependencies Through Formal Composition Rules} The modeling tool also allows the functionality and resource interfaces of DPIs to be connected through formal composition rules. Specifically, a functionality produced by one DPI can become the resource consumed by another. These compositions explicitly preserve the dependencies among subsystems, making the propagation of requirements across the system easier to trace. 

\subsubsection*{Addressing Challenge 3---Separation of Feasibility Analysis and Final Design Selection} The analysis tool characterizes the feasible design space for the overall system and identifies the Pareto-optimal design alternatives for a required system functionality. However, it does not prescribe how to select a final design from this set. That choice depends on the designer’s preference or application-specific objective. This contrasts with optimization-driven co-design approaches, in which feasibility analysis and design selection are typically handled within the same formulation.
%\textcolor{blue}{Suggestion: The analysis tool characterizes the feasible design space for the overall system and identifies the Pareto-optimal design alternatives for a required system functionality. It does not choose one final design, because that choice depends on the designer's preference or application-specific objective. In contrast, conventional optimization-driven co-design typically combines feasibility analysis and final design selection within the same optimization problem.}
%
%Compositional analysis identifies a set of Pareto-optimal design alternatives for the overall system given a required functionality, but a final design must still be selected from this set. This
\subsection*{Discussion: Complementarity of the Two Co-Design Approaches}
The separation between feasibility analysis and final design selection reveals a natural complementarity between composition-driven and optimization-driven co-design: composition-driven methods can characterize the feasible design space and identify Pareto-optimal alternatives, while optimization-driven methods can select a final design through a secondary optimization over the Pareto-optimal set. 

Applying composition before optimization may be particularly relevant to cyber-physical B6G architectures such as CPIE and ISCAC, which involve heterogeneous subsystems (e.g., communication, sensing, control, actuation, and computing) whose functionality and resource requirements propagate across the architecture. As these systems expand, explicitly representing such dependencies becomes increasingly valuable for characterizing the system-level design space.

The resulting separation enables a potential offline-online implementation. Computationally intensive compositional analysis can first be performed offline to characterize the system design space. Then, online optimization can select a final design from the resulting Pareto-optimal alternatives as system requirements and available resources change. Table~\ref{tab:design_comparison} compares \textit{optimization-driven} and \textit{composition-driven} co-design approaches across several analytical aspects.
%This restricts the online search to the Pareto-optimal set rather than the entire design space.   
%
\section*{Composition-Driven Co-Design Methodology}
In this section, we develop a composition-driven methodology for co-designing communication and control subsystems in cyber-physical B6G networks. To illustrate its stages, we present a running case study of a wireless-assisted control system for a single autonomous robot connected to a radio access network (RAN). For clarity of presentation, we intentionally adopt simplified subsystem models and design parameters for the case study. However, the methodology is independent of these assumptions and extends to more complex systems. In each stage, we first present the relevant theoretical elements of composition, followed by their application to the robot control solution. To be consistent with the nature of magazine articles, we limit mathematical notation to what we consider essential. A rigorous mathematical treatment of the theoretical elements is provided by Censi in~\cite{censi2016mathematicaltheorycodesign}. 

\subsection*{Stage I: Subsystem Abstraction and DPI Representation}
%as the system-level functionality requested by
%\subsubsection*{Theoretical Tool 1---Subsystem Abstraction and Composition} 
The first stage answers the question: \emph{What functionality does each subsystem provide, and what resources does it consume?} Each subsystem is represented independently through the DPI abstraction. A DPI is a mathematical structure consisting of two partially ordered sets: functionalities $\mathcal{F}$ and resources $\mathcal{R}$, together with a set of implementations $\mathcal{I}$. These sets are connected through two mappings: (i) $\operatorname{exec}:\mathcal{I}\rightarrow \mathcal{F}$ that captures the functionalities a given implementation delivers, and (ii) $\operatorname{eval}:\mathcal{I}\rightarrow\mathcal{R}$ that determines the resources a given implementation consumes. A DPI model $\langle\mathcal{F},\mathcal{R},\mathcal{I},\operatorname{exec},\operatorname{eval}\rangle$ induces a family of optimization problems of the form: \textit{given a required functionality $f \in \mathcal{F}$, find the implementation $i \in \mathcal{I}$ that delivers $f$ or higher while minimizing resource consumption $r =  \operatorname{eval}(i)$, $r \in \mathcal{R}$}. Formally:
\begin{equation}
    \min_{i \in \mathcal{I}} \; \operatorname{eval}(i) \quad \text{s.t.} \quad \operatorname{exec}(i) \geq f.
    \label{eq:queryDPI}
\end{equation}
Thus, a DPI establishes an explicit relationship between a required functionality and the minimum resources needed to provide it.
%Alternatively, \textit{given a resource budget, find the implementation that maximizes the achievable functionality subject to the available resources}. In this case, Eq.~(\ref{eq:queryDPI}) is modified by replacing $\min$ with $\max$ and interchanging the roles of $f$ and $r$.  
%\subsubsection*{Theoretical Tool 2---Characterization of Composed System Dependencies}

It is also useful to characterize a DPI as a feasibility map $h:\mathcal{F} \rightarrow \mathcal{AR}$, where $\mathcal{AR}$ is the set of antichains of $\mathcal{R}$.\footnote{In order theory, an antichain is a set of mutually non-dominated elements under a partial order.} For each required functionality $f$, $h(f)$ returns the solution set of the optimization problem in (\ref{eq:queryDPI}). Evaluating $h(f)$ answers two fundamental questions: (i) whether $f$ is feasible and, when feasible, (ii) which minimal (i.e., non-dominated) resource allocations are capable of achieving it. In other words, $h(f)$ is the Pareto front of resource allocations for $f$. If $h$ is monotone non-decreasing, then the DPI is called a monotone DPI (MDPI). Monotonicity guarantees that more demanding functionality cannot consume fewer resources, a behavior that is generally desired for engineering systems.%Monotonicity enables a meaningful notion of minimal resource allocations It prevents paradoxical functionality-resource relationships while enabling a meaningful notion of minimal resource allocations. 

\paragraph*{Running Case Study} Assume a robot unit (RU) moves along one dimension towards a desired reference position $r$. The RU periodically measures its state and communicates it through the RAN to an edge controller (K). K uses the RU state information to compute control actions. A management module (MM) at the network core interfaces the system with an external application that specifies the required functionality for the robot. The requested functionality consists of a maximum admissible position error $\epsilon$ and a maximum time $T_{\mathrm{max}}$ to achieve this accuracy, denoted by the tuple $(\epsilon,T_{\mathrm{max}})$.

We represent this architecture using four subsystem DPIs: RU, K, RAN, and MM DPIs. For each DPI, we identify the resources it consumes and the functionality it provides. The RU DPI consumes control actions and uplink data rate $R_{\mathrm{UL}}$ to provide robot state information. The K DPI consumes robot state information to provide control actions and the requested functionality $(\epsilon,T_{\mathrm{max}})$. We assume that state updates dominate the data volume of the 
control actions and therefore neglect the downlink communication requirement. The RAN DPI consumes uplink bandwidth $B$ and the robot's uplink transmission power $P$ to provide $R_{\mathrm{UL}}$. The MM DPI exposes the functionality $(\epsilon,T_{\mathrm{max}})$ provided by K to the external application. For readability, we henceforth refer to the subsystem DPIs simply as RU, K, RAN, and MM. At this stage, their feasibility maps $h_{\mathrm{RU}}$, $h_{\mathrm{K}}$, $h_{\mathrm{RAN}}$, and $h_{\mathrm{MM}}$ are assumed to be available and monotone. Their derivation from subsystem models is discussed in Stage III. 

\subsection*{Stage II: DPI Composition and System-level MCDPI Formation}
\begin{figure}[htbp]
\centering
\begin{subfigure}[b]{\columnwidth}
\centering
\includegraphics[height=1.6in]{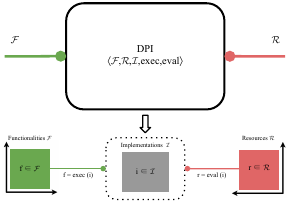}
\caption{Basic DPI model.}
\label{}
\end{subfigure}
\\
\vspace{0.5cm}
\begin{subfigure}[b]{\columnwidth}
\centering
\includegraphics[height=0.55in]{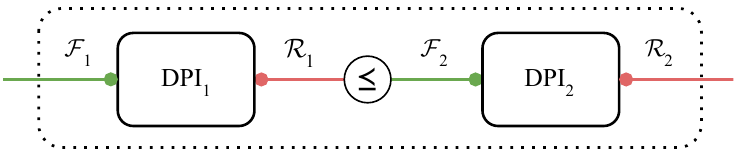}
\caption{Series composition $\mathrm{DPI_1} \bigodot \mathrm{DPI_2}$.}
\label{}
\end{subfigure}
\\
\vspace{0.5cm}
\begin{subfigure}[b]{\columnwidth}
\centering
\includegraphics[height=0.9in]{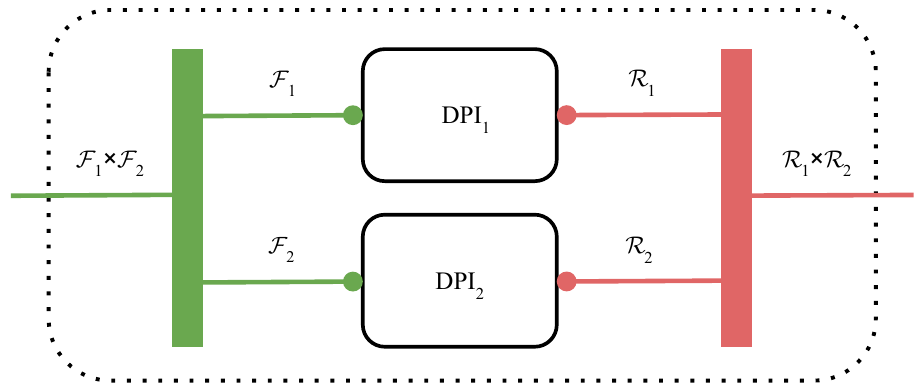}
\caption{Parallel composition $\mathrm{DPI_1} \bigotimes \mathrm{DPI_2}$.}
\label{}
\end{subfigure}
\\
\vspace{0.5cm}
\begin{subfigure}[b]{\columnwidth}
\centering
\includegraphics[height=0.9in]{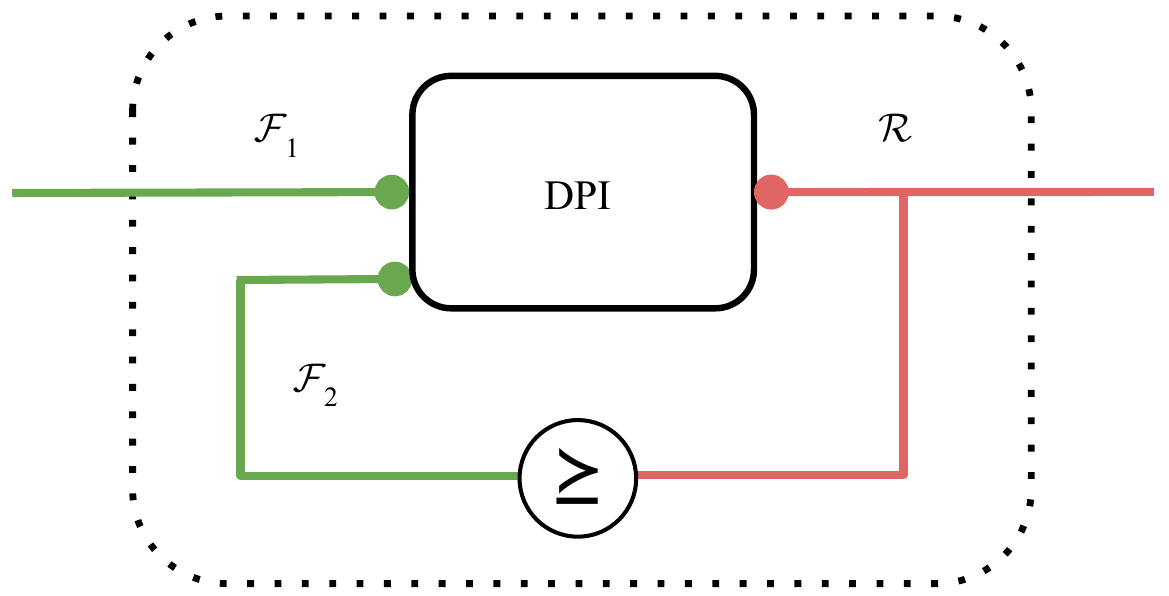}
\caption{Loop composition $\mathrm{DPI}^\dagger$.}
\label{}
\end{subfigure}
\\
\caption{Illustration of the basic DPI model and the three composition operators: series $\bigodot$, parallel $\bigotimes$, and loop $\dagger$. The symbols $\succeq$ and $\preceq$ denote interconnection between functionalities and resources, with $\mathcal{F} \succeq \mathcal{R}$~\cite{censi2016mathematicaltheorycodesign}. The compositions in (b), (c), and (d) form CDPIs.}
\label{fig:compositionOperators}
\end{figure}
The second stage answers the question:
\emph{How are the functionality and resource requirements of the individual subsystems interconnected?} A co-design problem with implementation (CDPI) is defined as the tuple $\langle \mathcal{F},\mathcal{R},\langle \mathcal{V},\mathcal{E}\rangle \rangle$, where $\mathcal{F}$ and $\mathcal{R}$ are the functionalities and resources of the overall system, and $\langle \mathcal{V},\mathcal{E} \rangle$ is a multigraph\footnote{A multigraph is a graph that allows multiple edges between the same pair of vertices.} of its constituent DPIs $\mathcal{V}$ and their interconnections $\mathcal{E}$. Interconnections capture functionality–resource dependencies among design problems. Specifically, each DPI may consume external resources as well as functionalities from other DPIs to produce its own functionality. A CDPI is obtained through the repeated application of three formal composition operators: series, parallel, and loop. %A brief explanation follows.%These dependencies motivate the notion of composition.
\begin{itemize}
    \item \textit{Series Composition}: Given two DPIs, the functionality provided by $\mathrm{DPI_1}$ is consumed by $\mathrm{DPI_2}$. Series composition is denoted by $\mathrm{DPI_1} \bigodot \mathrm{DPI_2}$. 
    \item \textit{Parallel Composition $ $}: $\mathrm{DPI_1}$ and $\mathrm{DPI_2}$ contribute their own functionality and consume their own resources. Parallel composition is denoted by $\mathrm{DPI_1} \bigotimes \mathrm{DPI_2}$. 
    \item \textit{Loop Composition$ $}: Given a $\mathrm{DPI}$, a factor of its provided functionality is also consumed as a resource, closing a loop. Loop composition is denoted by $\mathrm{DPI}^\dagger$. 
\end{itemize}
The formal theory of co-design guarantees that monotonicity is preserved under composition~\cite{censi2016mathematicaltheorycodesign}. Specifically, if all constituent DPIs are MDPIs, then the feasibility map of the composed system is also monotone, constituting a monotone CDPI (MCDPI). Figure~\ref{fig:compositionOperators} illustrates the basic DPI model and the composition operators. 
%that monotonicity is closed under composition, meaning that , then the resulting composition is also monotone, a monotone CDPI (MCDPI).

\begin{figure*}[t]
\centering
%\begin{adjustbox}{max width=\textwidth}%{Frame=1pt, Margin=10pt, center} % Keeps your full frame intact
\nopagebreak
\begin{minipage}[b]{0.40\textwidth}
\begin{subfigure}[b]{\textwidth}
\centering
\includegraphics[height=1.5in]{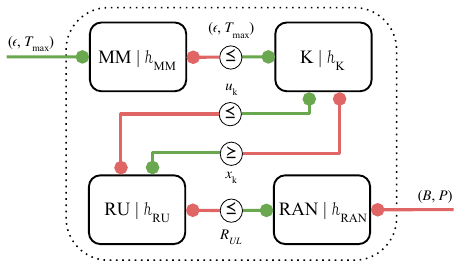}
\caption{Composition graph of RU, K, RAN, and MM.}
\label{fig:compositionGraph}
\end{subfigure}
\\
%\vspace{0.1cm}
\begin{subfigure}[b]{\textwidth}
\centering
\includegraphics[height=1in]{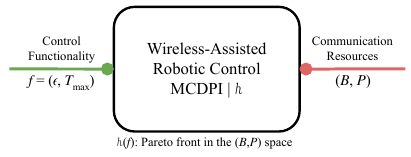}
\caption{Black-box view of the composed MCDPI and its feasibility map $h$.}
\label{fig:blackBoxMCDPI}
\end{subfigure}
\end{minipage}%
\hfill
\hfill
\begin{minipage}[b]{0.50\textwidth}
\begin{subfigure}[b]{0.9\textwidth}
\centering
\includegraphics[height=2.250in]{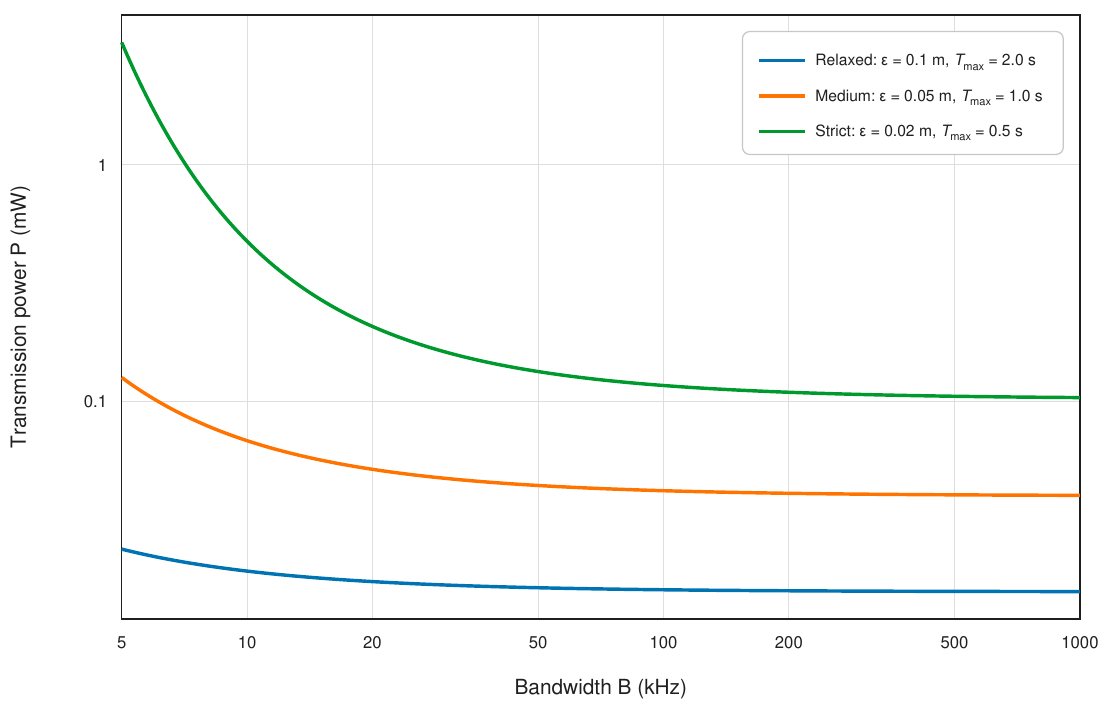}
\caption{Pareto fronts $h(f)$ in the $(B,P)$ space for different required control functionalities $f=(\epsilon,T_{\mathrm{max}})$: relaxed $(0.1~\textrm{m},2~\textrm{s})$, medium $(0.05 ~\textrm{m},1~\textrm{s})$, and strict $(0.02~\textrm{m},0.5~\textrm{s})$. Fixed system parameters: $e_0=1.0~\textrm{m}$, $\rho=0.8$, $L=1024~\textrm{bits}$, $g_{\mathrm{dB}}=-120~\textrm{dB}$, and $N_0=4\times10^{-21}~\textrm{W/Hz}$.}
\label{fig:paretoFronts}
\end{subfigure}
\end{minipage}\\
\caption{Wireless-assisted robotic control case study. The composed architecture consumes external communication resources $(B,P)$ and exposes the control functionality $(\epsilon,T_{\textrm{max}})$. The results show the expected monotonic behavior of an MCDPI: more demanding control functionality consumes more communication resources.}
\label{fig:caseStudy}
%\end{adjustbox}
\end{figure*}

\paragraph*{Running Case Study} We compose RU, K, RAN, and MM according to their functionality–resource dependencies. The uplink data rate provided by RAN is consumed by RU. This dependency is represented through series composition. RU and K exhibit mutual dependencies, the robot state information provided by RU is consumed by K, and the control actions generated by K are consumed by RU. RU and K are then interconnected through a loop composition. Finally, MM consumes the functionality provided by K through series composition. Since these DPIs are assumed to be monotone, the resulting composition forms an MCDPI with feasibility map $h$. We can now hide the internal structure of the MCDPI and treat it as a black box design that consumes external communication resources $(B,P)$ and produces the control functionality $(\epsilon,T_{\mathrm{max}})$. The composition graph is illustrated in Figure~\ref{fig:compositionGraph} and the corresponding black-box view in Figure~\ref{fig:blackBoxMCDPI}. We emphasize that the graph represents the functionality–resource dependencies among the DPIs and should not be confused with a signal-flow diagram.

\subsection*{Stage III: MCDPI Feasibility and Pareto Analysis}
The third stage answers the question:
\emph{Can the requested system functionality be achieved, and which resource allocations can realize it?} Suppose we are given an MCDPI model $\langle \mathcal{F}, \mathcal{R}, \langle \mathcal{V},\mathcal{E} \rangle \rangle$ obtained through composition, for which the feasibility maps $h_v$ of the constituent MDPIs are available. Then, for a required functionality $f$, the formal theory of co-design provides three important results:
\begin{enumerate}
%     \item \textit{}. The least fixed point is the smallest such value with respect to the underlying partial order.
     \item \textit{The feasibility map $h$ of the MCDPI admits an explicit expression in terms of the maps $h_v$ and the operators defining its composition (i.e., $\bigodot$, $\bigotimes$, and $\dagger$).}
     \item \textit{If the MCDPI is formed exclusively through series and parallel composition, and thus contains no feedback loops, the Pareto front $h(f)$ is obtained by directly evaluating the expression for $h$ at $f$.} 
     \item \textit{If the MCDPI is formed using at least one loop composition, and thus contains feedback loops, the Pareto front $h(f)$ is given by the least fixed point of a monotone map\footnote{A fixed point of a map is a value that remains unchanged when the map is applied.} induced by $h$ over the set of antichains. Kleene's fixed-point algorithm can be used to find this solution, with computational complexity dependent on the number of feedback loops in the composed problem.}%
\end{enumerate}
%the feasibility maps of the constituent DPIs and the operators defining their composition.
\paragraph*{Running Case Study} We now derive the feasibility map $h$ for the wireless-assisted robotic control MCDPI formed in Stage II. Recall that, for a required functionality $f=(\epsilon,T_{\mathrm{max}})$, $h(f)$ determines (i) whether $f$ is feasible and, if so, (ii) the Pareto-optimal $(B,P)$ allocations that can achieve it. From result 1, the map $h$ admits an explicit expression in terms of $h_{\mathrm{RU}}$, $h_{\mathrm{K}}$, $h_{\mathrm{RAN}}$, and $h_{\mathrm{MM}}$. Based on the composition graph in~Figure \ref{fig:compositionGraph}, this expression is given by
\begin{equation}
\small
h= h_\mathrm{RAN} \bigodot \left(h_{\mathrm{RU}} \bigotimes h_\mathrm{K} \right)^{\dagger} \bigodot h_{\mathrm{MM}},
\label{eq:h_mcdpi}
\end{equation}
%$h_{\mathrm{RU}}$, $h_{\mathrm{K}}$, $h_{\mathrm{RAN}}$, and $h_{\mathrm{MM}}$
where $\bigodot$, $\bigotimes$ and $\dagger$ denote series, parallel, and loop composition, respectively. We characterize the subsystem $h$'s next using simplified subsystem models. 
%Evaluating (\ref{eq:h_mcdpi}) requires expressions for . 

For RU, we consider a discrete-time robot model $x_{k+1}=x_k+T_su_k$, where $x_k$ is the robot position at the $k$-th time step and $u_k$ is a velocity command the robot consumes from K. The robot position is sampled every $T_s$ seconds and encoded into a fixed $L$-bit representation for uplink transmission. RU therefore sends state updates to K at a rate $\nu_s=1/T_s$, requiring an uplink data rate $R_{\mathrm{UL}} \geq L\nu_s$. Then, $h_{\mathrm{RU}}$ gives the minimum uplink data rate $R_{\mathrm{UL}}$ required for a given $\nu_s$.

For K, we consider a proportional feedback controller that generates the velocity commands $u_k = K_p(r-x_k)$, where $K_p>0$ is the controller gain and $e_k=r-x_k$ is the position error at the $k$-th time step. Combined with the robot dynamics, the error evolves as $e_{k+1}=(1-T_sK_p)e_k$. We set $K_p=(1-\rho)\nu_s$ so that $0<\rho<1$ determines how much the position error is reduced after each state update. Hence, $e_k=\rho^ke_0$. We denote by $N$ the minimum $k$ for which $|e_k| \leq \epsilon$. Given $T_{\mathrm{max}}$, these updates must satisfy $NT_s \leq T_{\mathrm{max}}$. Smaller values of $\epsilon$ or $T_{\mathrm{max}}$ therefore require smaller $T_s$, and consequently a higher state-update rate $\nu_s$. Thus, $h_{\mathrm{K}}$ gives the minimum $\nu_s$ required for a given $(\epsilon,T_{\mathrm{max}})$.

For RAN, we assume an AWGN channel and use the Shannon capacity model $R_{\mathrm{UL}}=B\log_2\!\left(1+\frac{Pg}{N_0B}\right)$ to relate $R_{\mathrm{UL}}$ to $B$ and $P$, where $g$ is the channel gain and $N_0$ is the noise power spectral density. Therefore, $h_{\mathrm{RAN}}$ gives the Pareto-optimal $(B,P)$ allocations for a given $R_{\mathrm{UL}}$. For MM, $h_{\mathrm{MM}}$ is trivial. MM consumes $(\epsilon,T_{\mathrm{max}})$ from K and simply exposes it to the external application, without requiring additional resources. 

With all subsystem $h$'s defined, the composition expression in~(\ref{eq:h_mcdpi}) can be evaluated using the Python co-design library in~\cite{briat2026textttcodesignmcdppythonlibrarymonotone}. Figure~\ref{fig:paretoFronts} illustrates the resulting Pareto fronts $h(f)$ in the $(B,P)$ space for different control functionalities $f=(\epsilon,T_{\mathrm{max}})$. The results exhibit the expected monotonic behavior of an MCDPI: as the required control functionality becomes more demanding, the system consumes more communication resources, shifting the corresponding Pareto fronts upwards. 
%The library implements provides a Kleene fixed-point solver to handle feedback loops when necessary. In our simplified case study, however, the RU-K loop term $\left(h_{\mathrm{RU}} \bigotimes h_{\mathrm{K}}\right)^\dagger$ is handled analytically. 
%Specifically, combining the RU and K models eliminates the internal feedback variables $u_k$ and $x_k$. This reduces the loop term to a direct mapping from the required functionality $(\epsilon,T_{\mathrm{max}})$ to the uplink data rate $R_{\mathrm{UL}}$. Therefore, evaluating the Pareto front does not require fixed-point computation. The analysis of systems with more complex feedback dependencies is left for future work. 
%requiring fixed-point computation
\subsection*{Stage IV: Pareto-Informed Design Selection}
The fourth stage answers the question:
\emph{Which design should be selected from the feasible Pareto-optimal alternatives?} The compositional analysis identifies the feasible and non-dominated design alternatives, but it does not impose a particular preference among them. A final design can therefore be selected by introducing application-specific objectives and constraints over the Pareto set obtained in Stage III. In general, this can be written as
\(r^\star =\arg\min_{r\in h(f)} J(r)\), where $J(r)$ represents the design preference.
%Given the Pareto-optimal alternatives obtained in Stage III, a final design is selected from this set. 
\paragraph*{Running Case Study} 
For the robotic control system, we assume that the uplink transmission power $P$ of the robot is limited by a budget $P_{\textrm{max}}$. Then, a final design can be selected by minimizing bandwidth: \(\min_{(B,P)\in h(f)} B \quad \text{s.t.}\quad P\leq P_{\max}\). For $P_{\textrm{max}}=0.1~\textrm{mW}$, the strict control requirement $f=(0.02~\textrm{m},0.5~\textrm{s})$ is simply infeasible regardless of bandwidth. The medium and relaxed control requirements are feasible, requiring minimum bandwidths of approximately $6.1~\textrm{kHz}$ and $1.3~\textrm{kHz}$, respectively. As noted earlier, the compositional analysis does not prescribe how to pick a final design. However, the final selection is naturally Pareto-informed. 
%This stage is important because the compositional analysis from Stages I, II, and III reveals design alternatives for the system but does not prescribe how to select a final design among them. However, the final selection is naturally Pareto-informed, involving additional decision criteria not considered in the preceding compositional analysis.
%Then, design alternatives are further restricted to those satisfying $P \leq P_{\textrm{max}}$. Finally, a final design is selected based on minimum bandwidth usage. 
%
\section*{Research Challenges and Future Directions}
The practical adoption of a composition-driven co-design methodology presents several research challenges. Key directions for addressing them are discussed next.
%While composition-driven co-design offers an intuitive framework for reasoning about complex B6G networks, putting it into practice 
%
\subsection*{Realistic DPI Instantiations}
The quality of the compositional analysis depends on accurate DPI instantiations\footnote{A DPI instantiation specifies the tuple $\langle\mathcal{F},\mathcal{R},\mathcal{I},\operatorname{exec},\operatorname{eval}\rangle$ for a subsystem.} that represent subsystem behavior. In cyber-physical B6G, this requires the DPI mappings (i.e., $\operatorname{exec}$ and $\operatorname{eval}$) to embed analytical, data-driven, or hybrid models of communication, sensing, control, actuation and computing subsystems. The key challenge is to incorporate these models while preserving DPI monotonicity, which is central to compositional analysis. This creates research opportunities for constructing realistic and monotone DPI instances by leveraging state-of-the-art developments in wireless channel modeling, perception and localization, learning-based control, and cloud/edge/fog/dust computing.
%
%The analysis of large composed systems can be computationally challenging. Kleene's fixed-point algorithm may not scale well to  large B6G architectures with numerous subsystem dependencies and feedback loops.
\subsection*{Scalable MCDPI Analysis}
As B6G systems incorporate increasing numbers of interacting subsystems, future research should investigate scalable methods for MCDPI analysis, including computationally efficient algorithms for the exact evaluation or approximation of $h(f)$ and hierarchical system composition strategies that can reduce the number of feedback loops while preserving structural subsystem dependencies. Another promising direction is distributed MCDPI analysis, in which network entities collectively construct and maintain a global Pareto-optimal design space rather than relying on a centralized entity with a complete view of the system. 

\subsection*{Advances in Pareto-Informed Design Selection} 
The complementarity between composition-driven and optimization-driven co-design creates research opportunities for Pareto-informed design selection. An important research direction is the development of decision methods that efficiently navigate Pareto-optimal design alternatives. These methods may include optimization strategies that support adaptive design selection by navigating both within the Pareto front associated with a required functionality and across Pareto fronts associated with different functionality requirements as the required functionality, available resources, or additional design objectives change. 

\section*{Conclusion}
Future B6G networks are expected to evolve from communication-centric infrastructures into complex cyber-physical systems in which communication and control subsystems are mutually dependent. In this article, we explored how the formal theory of co-design can provide a compositional perspective for the co-design of communication and control in cyber-physical B6G. Although the compositional co-design perspective is distinct from the conventional optimization-driven approach, we view the two as complementary rather than competing. Composition-driven co-design can characterize Pareto-optimal design alternatives. Optimization-driven methods can then perform the final design selection through a secondary optimization over this Pareto-optimal set.

%We hope this perspective inspires further research at the intersection of the formal theory of co-design and optimization methods, leading to more systematic co-design methodologies for cyber-physical B6G networks.

%\textcolor{blue}{We hope that this perspective will motivate further research at the intersection of formal co-design theory and optimization, ultimately enabling more systematic and scalable co-design methodologies for cyber-physical B6G networks.}
%Unlike the optimization-driven approach, composition-driven co-design is guided by modular abstractions and formal composition rules. 
%Unlike the optimization-driven approach that rely entirely on optimization, composition-driven co-design abstracts subsystems as modular design problems, composes them through formal composition rules, and characterizes the functionality–resource dependencies and associated design tradeoffs in the resulting system. 
%these interactions. 
%helps designers think of complex systems in terms of the assemble of 
%The theory formalizes the intuitive engineering practice of designing complex systems by assembling simpler subsystems.
%does not rely on the direct optimization of particular objectives. Instead, it 
%We argue that this compositional perspective guided by modular abstractions and formal composition rules is particularly well suited for future cyber-physical B6G networks, whose architectures are expected to become increasingly heterogeneous and highly interconnected. 
%
\bibliographystyle{IEEEtran}
\bibliography{references}
\begingroup
\let\description\LaTeXdescription
\let\enddescription\endLaTeXdescription
%\printacronyms  
\endgroup
%

%\begin{IEEEbiographynophoto}{Angelo Vera-Rivera}(Member, IEEE) is a Ph.D. student in the Department of Electrical and Computer Engineering, University of Manitoba, Canada.
%\end{IEEEbiographynophoto}
%
%\begin{IEEEbiographynophoto}{Ekram Hossain}(Fellow, IEEE) is a Professor in the Department of Electrical and Computer Engineering, University of Manitoba, Canada.
%\end{IEEEbiographynophoto}
%
%\begin{IEEEbiographynophoto}{Atefeh Termehchi} is a Research Associate in the Department of Electrical and Computer Engineering, University of Manitoba, Canada.
%\end{IEEEbiographynophoto}
%
\end{document}